\newcommand{\idty}{{\leavevmode{\rm 1\mkern -5.4mu I}}}
\newcommand{\Ir}{Z\!\!\!Z}
\newcommand{\Ibb}[1]{ {\rm I\ifmmode\mkern
            -3.6mu\else\kern -.2em\fi#1}}
\newcommand{\ibb}[1]{\leavevmode\hbox{\kern.3em\vrule
     height 1.2ex depth -.3ex width .2pt\kern-.3em\rm#1}}
\newcommand{\Cx}{{\ibb C}}

\newcommand{\Nl}{{\Ibb N}}
\documentstyle[preprint,aps]{revtex}
\tighten
\begin{document}
\draft
\preprint{GOET-TP 88/94}
\title{\huge Discrete Differential Manifolds  \\
        and Dynamics on Networks}
\vskip1.5cm
\author{\bf Aristophanes Dimakis \vskip.2cm}
\vskip.4cm
\address{Department of Mathematics, University of Crete,
      GR-71409 Iraklion, Greece}
\author{\bf Folkert M\"uller-Hoissen \vskip.2cm}
\vskip.4cm
\address{Institut f\"ur Theoretische Physik,
      Bunsenstr. 9, D-37073 G\"ottingen, Germany}
\author{\bf Francois Vanderseypen$\natural$ \vskip.2cm}
\address{Instituut voor Theoretische Fysica, Katholieke Universiteit,
         B-3001 Leuven, Belgium \\
         $\natural$ Aspirant NFWO}

\maketitle

\begin{abstract}
\noindent
A {\em discrete differential manifold} we call a countable set
together with an algebraic differential calculus on it. This
structure has already been explored in previous work and
provides us with a convenient framework for the formulation of
dynamical models on networks and physical theories with discrete space
and time. We present several examples and introduce a notion of
differentiability of maps between discrete differential manifolds.
Particular attention is given to differentiable curves in such spaces.
Every discrete differentiable manifold carries a topology and we
show that differentiability of a map implies continuity.
\end{abstract}

\pacs{02.10.+w, 02.40.+m, 03.20.+i, 04.20.Cv}

\renewcommand{\theequation} {\arabic{section}.\arabic{equation}}

\section{Introduction}
\setcounter{equation}{0}
The aim of our present work is to develop a mathematical formalism
which allows an {\em intrinsic} formulation of dynamics and field
theory on networks. By a network we mean a directed graph (digraph)
which consists of a set of points (vertices) and a set of arrows
connecting pairs of points. The formalism should give us a natural way
to guarantee, for example, that a `particle' hopping in discrete time
steps on the set of vertices of a network respects the network structure
in the sense that the motion can only take place in the direction of
existing arrows of the digraph.
\vskip.3cm

In \cite{DMH94} it was found that a digraph with at most two
antiparallel arrows between each pair of vertices determines an
`algebraic differential calculus' on the set $\cal M$ of vertices of
the digraph (respectively, on the algebra of functions on
this set). This observation was crucial for our solution of the
problem mentioned above.
\vskip.3cm

An algebraic differential calculus is an analogue of the calculus of
differential forms on a manifold. It should be regarded as a basic
structure for the formulation of dynamical systems and field theories.
In the present case, the differential
calculus is `noncommutative' in the sense that differential forms and
functions do not commute, in general. It fits into the more general
framework of noncommutative geometry (see \cite{Conn86}). Indeed,
some of the constructions used in this work are defined on arbitrary
(not necessarily commutative) associative algebras. In this sense our
choice of the commutative algebra of functions on a discrete set is
just an example. But in the latter case we are able to associate a
physical picture with the formalism. A comparable understanding is
lacking in the case of noncommutative algebras.
\vskip.3cm

The physical motivations for our work are `manifold'. In particular,
ideas about a discrete space (-time) structure and space-time as a
network stimulated our interest. There is already a vast literature in
this field, but especially close to our work seems to be
\cite{Sork91,Bomb87,Zapa93}.
In the present work we are not going beyond the classical level.
Ideas about quantization of topology and space-time can be pursued
in this framework (see also \cite{DMH94}) and we should then expect
relations, e.g., with the work by Isham on `quantum topology'
\cite{Isha89} and Finkelstein's work on `quantum space-time
networks' \cite{Fink89}. In addition we should mention the work
on `pregeometry' in the sense of \cite{Anto94} and references given
there. Of special interest is also 't Hooft's
work \cite{tHoo88} suggesting that a theory which at large distance
scales behaves like a quantum field theory may be deterministic and
discrete at a small length scale (which may be the Planck scale).
\vskip.3cm

The usefulness of algebraic differential calculus was demonstrated in
\cite{DMHS93} for lattice field theories. More generally, differential
calculus on discrete sets was then developed in \cite{DMH94-fin,DMH94}
(see also \cite{Sita92} for the case of discrete groups). A special
example is the 2-point space which appeared in particle physics models
of noncommutative geometry \cite{Conn+Lott90}.
\vskip.3cm

We shall now make more precise with what kind of mathematical
framework we are working.
Let $\cal M$ be a discrete (in the sense of countable) set. An algebraic
{\em differential calculus} on $\cal M$ is an extension of the algebra
$\cal A$ of $\Cx$-valued functions on $\cal M$ to a differential algebra
$(\Omega({\cal M}),d)$. Here $\Omega({\cal M}) = \bigoplus_{r=0}^\infty
\Omega^r({\cal M})$ is a $\Ir$-graded associative algebra where
$\Omega^0({\cal M}) = \cal A$ and $\Omega^{r+1}({\cal M})$ is
generated as an $\cal A$-bimodule via the action of a linear operator
$d \, : \, \Omega^r({\cal M}) \rightarrow \Omega^{r+1}({\cal M})$.
This operator is assumed to satisfy $ d^2=0$ and the graded Leibniz rule
$ d(\omega \omega') = (d\omega) \, \omega'+(-1)^r \omega \, d\omega' $
where $\omega \in \Omega^r({\cal M})$. It has been shown in
\cite{DMH94} how this structure supplies $\cal M$ with a topology
and assigns a (local) notion of dimension to it. Each differential
algebra on $\cal M$ can be obtained as the quotient of the so-called
universal differential algebra by some differential ideal. A
systematic construction of such `reductions' of the universal
differential algebra has been given in \cite{DMH94}.
\vskip.3cm
\noindent
We take the point of view that a discrete set $\cal M$ supplied with
a differential calculus may be regarded as a kind of analogue of a
(continuous) differentiable manifold. This is justified by the
results in \cite{DMH94} and suggests the following definition.
\vskip.3cm
\noindent
{\bf Definition.} A {\em discrete differential manifold} is a
discrete set $\cal M$ together with a differential calculus on it.
\vskip.3cm
\noindent
This is the basic structure which we explore in the following.
After recalling differential calculus on discrete sets in section II
we collect some examples in section III. A notion of a `differentiable
map' between discrete differential manifolds is the subject of
section IV. Special cases are `diffeomorphisms' (section V) and
`differentiable curves' (section VI). A natural further step is to
consider the set of all differentiable curves in a given
discrete differential manifold (section VII). Every discrete
differential manifold carries the structure of a topological space.
More precisely, it determines a larger set $\hat{M}$ with a
topology on it. This is the subject of section VIII. In section IX we
show that differentiability of a map implies continuity.
Section X contains some conclusions and additional remarks.
\vskip.3cm
\noindent
The formalism as presented in this paper basically applies to the case
of a finite set. For an infinite set some of the calculations are formal
and more efforts have to be invested to put things on a rigorous
footing.

\section{Differential calculus on discrete sets}
\setcounter{equation}{0}
With each element $i \in {\cal M}$ we associate a function $e_i \in
{\cal A}$ via
\begin{eqnarray}
      e_i(j) = \delta_{ij}      \; .
\end{eqnarray}
Then
\begin{eqnarray}                    \label{e-rels}
   e_i \, e_j = \delta_{ij} \, e_j  \qquad  \sum_i e_i = \idty
\end{eqnarray}
where $\idty(i) = 1 \; \forall i \in {\cal M}$. Acting with $d$ on
these relations and using the Leibniz rule yields
\begin{eqnarray}                  \label{ede}
  e_i \, d e_j = - d e_i \; e_j + \delta_{ij} \, d e_j
              \qquad \qquad
  \sum_i d e_i = 0    \; .
\end{eqnarray}
The special 1-forms
\begin{eqnarray}
     e_{ij} := e_i \, d e_j    \qquad  (i \neq j)
\end{eqnarray}
($e_{ii} := 0$) satisfy
\begin{eqnarray}       \label{e-e}
   e_i \, e_{jk} = \delta_{ij} \, e_{jk} \qquad
   e_{ij} \, e_k = \delta_{jk} \, e_{ij} \qquad
   e_{ij} \, e_{k \ell} = \delta_{jk} \, e_{ij} e_{j \ell} \; .
\end{eqnarray}
As products of these 1-forms only the $(r-1)$-forms
\begin{eqnarray}            \label{e...}
  e_{i_1 \ldots i_r} := e_{i_1 i_2} \, e_{i_2 i_3} \cdots
                        e_{i_{r-1} i_r}
\end{eqnarray}
are therefore allowed not to vanish. On these forms the operator $d$
acts as follows,
\begin{eqnarray}
  de_{i_1 \ldots i_r}
  & = &  \sum_j\sum_{k=1}^{r+1} (-1)^{k+1} e_{i_1 \ldots i_{k-1}j i_k
         \ldots i_r}    \; .         \label{de...}
\end{eqnarray}
The 1-form
\begin{eqnarray}
                p := \sum_{k,\ell} e_{k \ell}
\end{eqnarray}
satisfies
\begin{eqnarray}
  p^r = \sum_{i_1, \ldots, i_{r+1}} e_{i_1 \ldots i_{r+1}}  \; .
\end{eqnarray}
For $r=1, \ldots, 4$ the formula (\ref{de...}) can be rewritten as
\begin{eqnarray}
  d e_i &=& \lbrack p , e_i \rbrack            \\
  d e_{ij} &=& \lbrace p , e_{ij} \rbrace - e_i \, p^2 \, e_j    \\
  d e_{ijk} &=& \lbrack p , e_{ijk} \rbrack + e_i \, ( p \, e_j \, p^2
              - p^2 \, e_j \, p ) \, e_k
\end{eqnarray}
and so forth. These formulae are easily obtained by using
$e_{ij} = e_i \, p \, e_j$ and
\begin{eqnarray}
                d p = p^2 + \sum_i e_i \, p^2 e_i  \; .
\end{eqnarray}
They display the deviation of $d$ from a graded commutator.
If no further relations are imposed, we are dealing with
the universal differential calculus which we denote as
$\tilde{\Omega}({\cal M})$. In this case the $e_{i_1 \ldots i_r}$
constitute a basis over $\Cx$ of $\tilde{\Omega}^{r-1}({\cal M})$ for
$r>1$.

\vskip.3cm
\noindent
A systematic way of constructing smaller differential algebras
from the universal one is given by setting linear combinations of
`basic' forms to zero (`{\em reduction}'). Setting a linear
combination of the $e_{i_1 \ldots i_{r+1}}$ to zero does
not influence those $r$-forms which do not appear in this equation
and also not the forms with grade $< r$. It leads, however, to
constraints for forms of grade $> r$ via the action of $d$. For
1-forms, the vanishing of a linear combination implies the vanishing
of each basic 1-form which appears in the sum. We will be mainly
concerned with reductions on the level of 1-forms.

\vskip.3cm
\noindent
It is convenient to associate a diagram with a differential
calculus on $\cal M$ as follows. On horizontal levels we draw
vertices corresponding to all the basic $r$-forms
$e_{i_1 \ldots i_{i_{r+1}}} \neq 0$ in such a way that vertices
representing $(r+1)$-forms are below those representing $r$-forms
($r \geq 0$). An arrow is drawn between two vertices on neighboring
levels if the corresponding basic forms appear in (\ref{de...}) whereby
the relative sign determines the orientation of the arrow. The result is
an {\em oriented Hasse diagram} which completely specifies the
differential calculus. Several examples can be found in \cite{DMH94}.
If in the differential calculus a linear combination of basic $r$-forms
($r>1$) vanishes, this determines one of the $r$-forms in terms of the
others. That one should then be discarded in the diagram. But one has
to add a corresponding note to the diagram in order to be able to
reconstruct the differential calculus from the diagram.

\vskip.3cm
\noindent
When we replace arrows by edges we
obtain a {\em Hasse diagram} which determines a topology on $\cal M$ or,
more precisely, on a certain extension of $\cal M$ (cf also
\cite{Sork91}). More details are given in section VIII.

\vskip.3cm
\noindent
We also associate a {\em digraph} with a differential calculus on
$\cal M$ in the following way. If $e_{ij} \neq 0$ we draw an arrow
from $i$ to $j$. If reductions are only considered on the level
of 1-forms, this digraph already contains all the information about
the differential calculus and the oriented Hasse diagram can be derived
from it (cf \cite{DMH94}).

\vskip.3cm
\noindent
A homomorphism of differential algebras $\Omega({\cal M}) \rightarrow
\Omega({\cal N})$ is an algebra homomorphism which intertwines the
respective $d$'s.
According to a general result (see \cite{Kast88}, Corollary 1.9), each
differential algebra $\Omega({\cal M})$ is the image of a homomorphism
of differential algebras $\pi \, : \, \tilde{\Omega}({\cal M})
\rightarrow \Omega({\cal M})$ where $\tilde{\Omega}({\cal M})$ is the
universal differential algebra on $\cal M$. It is therefore the quotient
$\Omega({\cal M}) = \tilde{\Omega}({\cal M})/{\cal I}$ by some two-sided
differential ideal $\cal I$ (the kernel of the homomorphism) in
$\tilde{\Omega}({\cal M})$. A `differential ideal' is an ideal which is
mapped by $d$ into itself.
This alternative description of differential calculi on $\cal M$ will
be helpful in the following sections.
The ideal $\cal I$ is generated by those linear combinations of basic
forms viewed as elements of $\tilde{\Omega}({\cal M})$ which vanish
in the reduced differential calculus $\Omega({\cal M})$.
\vskip.3cm
\noindent
{\bf Remark.} We have seen that there are different
differential calculi on $\cal M$ and thus different $d$'s. As a
consequence, $e_{ij}$ also depends on the choice of the calculus. For
the sake of notational simplicity we do not indicate this dependence
in the hope that the latter will be clear from the respective context
in which these symbols appear.
                                      \hfill  {\Large $\Box$}

\section{Examples of discrete differential manifolds}
\setcounter{equation}{0}
In this section we collect some examples of discrete differential
manifolds in the sense of the definition given in the introduction.
Examples 2-4 are taken from \cite{DMH94,DMHS93} where the reader
can find ample discussions. Here we concentrate on those formulae
which are needed in particular in section VI. Example 5 is
new and therefore presented in some more detail. We also point out
some ways to construct discrete differential manifolds from given
ones.

\vskip.3cm
\noindent
{\bf Example 1.} Let $\cal M$ be a discrete set. If we regard it as
a subset of $\Ir^n$, then
\begin{eqnarray}                 \label{x^mu}
   x^\mu := \sum_{a \in {\cal M}} a^\mu \, e_a
            \qquad (\mu =1, \ldots, n)
\end{eqnarray}
are natural coordinate functions on $\cal M$. With the help
of (\ref{ede}) and (\ref{de...}) one obtains
\begin{eqnarray}
 \lbrack d x^\mu , x^\nu \rbrack = \tau^{\mu \nu}
 \quad \mbox{where} \quad
 \tau^{\mu \nu} = \sum_{a,b} (a^\mu - b^\mu)(a^\nu - b^\nu) \, e_{ab}
 \; .
\end{eqnarray}
Furthermore, one finds
\begin{eqnarray}
 \lbrack \tau^{\mu \nu} , x^\lambda \rbrack =
   \sum_{a,b} (a^\mu - b^\mu)(a^\nu - b^\nu)(a^\lambda - b^\lambda)
   \, e_{ab} \; .
\end{eqnarray}
So far we did not specify the differential calculus. This will be done
in the following examples.
\hspace*{2cm}                                  \hfill  {\Large $\Box$}

\vskip.3cm
\noindent
{\bf Example 2.} Let ${\cal M} = \Ir^n$ and $\Omega({\cal M})$ the
differential calculus determined by
\begin{eqnarray}
    e_{ab} \neq 0 \quad \Leftrightarrow \quad
           b = a + \hat{\mu}  \quad \mbox{for some } \mu
\end{eqnarray}
where $\hat{\mu} = (\hat{\mu}^\nu) := (\delta^\nu_\mu)$. This is the
(oriented) lattice calculus first considered in \cite{DMHS93} (see also
\cite{DMH94}). We obtain in this case
\begin{eqnarray}
   \tau^{\mu \nu} = \delta^{\mu \nu} \, \sum_a e_{a,a+\hat{\mu}}
                  = \delta^{\mu \nu} \, d x^\mu   \; .
\end{eqnarray}
In terms of the coordinate functions (\ref{x^mu}) the reduction
condition thus reads
\begin{eqnarray}
 \lbrack d x^\mu , x^\nu \rbrack = \delta^{\mu \nu} \; d x^\nu   \; .
\end{eqnarray}
We refer to \cite{DMHS93} for applications of this calculus to
lattice field theories.
                                       \hfill  {\Large $\Box$}

\vskip.3cm
\noindent
{\bf Example 3.} Let ${\cal M} = \Ir$ and $\Omega({\cal M})$ be the
differential calculus on $\cal M$ determined by the condition
\begin{eqnarray}
               \lbrack d t , t \rbrack = d t           \label{t-dt}
\end{eqnarray}
in terms of the natural coordinate function $t(k) = k \in \Ir$. This
is a special case ($n=1$) of the previous example and corresponds to
the reduction
\begin{eqnarray}
      e_{ij} \neq 0 \quad \Leftrightarrow \quad  j = i + 1
\end{eqnarray}
of the universal calculus (see also \cite{DMHS93,DMH94}).
It assigns a 1-dimensional structure to $\Ir$. Now
\begin{eqnarray}
   d f(t) = d t \; (\partial_+ f)(t) = (\partial_- f)(t) \, d t
\end{eqnarray}
define functions $\partial_\pm f$ on $\cal M$. A simple calculation
(cf \cite{DMHS93}) shows that
\begin{eqnarray}             \label{partial+-}
    (\partial_+ f)(t) = f(t+1) - f(t)  \qquad
    (\partial_- f)(t) = f(t) - f(t-1)    \; .
\end{eqnarray}
We will also use the notation
\begin{eqnarray}
              \dot{f}(t) := f(t+1) - f(t)
\end{eqnarray}
instead of $(\partial_+ f)(t)$. For functions $f(t), h(t)$ the
relation (\ref{t-dt}) now generalizes to
\begin{eqnarray}                \label{df-h}
   \lbrack df(t), h(t) \rbrack
 = \dot{f}(t) \, \lbrack dt, h(t) \rbrack
 = \dot{f}(t) \, \lbrack dh(t) , t \rbrack
 = \dot{f}(t) \, \dot{h}(t) \, \lbrack dt, t \rbrack
 = \dot{f}(t) \, \dot{h}(t) \, dt \; .
\end{eqnarray}
We will take $(\Ir, \Omega({\cal M}))$ as a mathematical model
for the parameter space of discrete time. The notion of discrete time
in physics has been explored by many authors (see
\cite{Fink89,tHoo88,discrete_time}, for example). Instead of $\Ir$ we
may consider a subset of $\Ir$ like ${\bf N} := \lbrace 0, \ldots N-1
\rbrace$ with the induced differential calculus (a `submanifold' of
$(\Ir, \Omega({\cal M}))$, see below). In this case one has to pay
attention to the fact that $f \, dt = 0$ with a function
$f = \sum_{i=0}^{N-1} f(i) \, e_i$ does not imply that $f$ vanishes.
The equation does not determine $f(N-1)$. In the same way $dt \, f = 0$
leaves $f(0)$ undetermined.
                                      \hfill  {\Large $\Box$}

\vskip.3cm
\noindent
{\bf Example 4.} We choose ${\cal M} = \Ir^n$ with the reduction of
the universal calculus determined by the conditions
\begin{eqnarray}
 e_{ab} \neq 0 \quad \Leftrightarrow \quad
 b = a + \hat{\mu} \quad \mbox{or} \quad b = a-\hat{\mu}
 \quad \mbox{for some } \mu
\end{eqnarray}
where $\hat{\mu} = (\delta^\nu_\mu)$. This is the `symmetric lattice
calculus' discussed in \cite{DMH94}. One finds
\begin{eqnarray}
 \tau^{\mu \nu} = \delta^{\mu \nu} \, \tau^\mu
 \quad  \mbox{where} \quad
 \tau^\mu := \sum_{a, \epsilon = \pm 1}  e_{a, a+ \epsilon \hat{\mu}}
\end{eqnarray}
so that
\begin{eqnarray}        \label{xdx-symlatt}
 \lbrack d x^\mu , x^\nu \rbrack = \delta^{\mu \nu} \, \tau^\mu
 \qquad \quad
 \lbrack \tau^\mu , x^\nu \rbrack = \delta^{\mu \nu} \, d x^\nu \; .
\end{eqnarray}
                                       \hfill  {\Large $\Box$}

\vskip.3cm
\noindent
{\bf Example 5.}  Let ${\cal M} = \Ir_N$. A reduction of the universal
differential calculus on $\Ir_N$ is given by
\begin{eqnarray}
   e_{ij} \neq 0  \quad \Leftrightarrow \quad j = i + \epsilon
          \; \mbox{mod} \, N
\end{eqnarray}
where $\epsilon = \pm 1$.
The associated digraph assigns to $\Ir_N$ the structure of a closed
(i.e. periodic) lattice which is `symmetric' in the sense that any two
neighboring sites are connected by a pair of antiparallel arrows.
Let $q \in \Cx$ be a primitive $N$th root of unity, i.e. $q^N = 1$,
and define
\begin{eqnarray}
            y := \sum_{i=0}^{N-1} q^i \, e_i   \; .
\end{eqnarray}
Then
\begin{eqnarray}             \label{dy}
   d y = y \, \lbrack (q-1) \, e^+ + (q^{-1} -1) \, e^- \rbrack
\end{eqnarray}
where
\begin{eqnarray}
          e^\epsilon := \sum_k e_{k, k+\epsilon}   \; .
\end{eqnarray}
On the lhs, $\epsilon$ stands for $\pm$ (instead of $\pm 1$). Using
(\ref{e-e}) we find
\begin{eqnarray}
       \lbrack d y , y \rbrack
  = y^2 \, \sum_\epsilon (q^\epsilon - 1)^2 \, e^\epsilon
  = (q^{1/2} - q^{-1/2})^2 \, y \, dy + \tau           \label{dy-y}
\end{eqnarray}
with
\begin{eqnarray}                       \label{tau}
 \tau := (q^{1/2} - q^{-1/2})^2 \, y^2 \, \sum_\epsilon e^\epsilon \; .
\end{eqnarray}
Furthermore,
\begin{eqnarray}                       \label{tau-y}
 \lbrack \tau , y \rbrack = (q^{1/2} - q^{-1/2})^2 \, y^2 \, dy   \; .
\end{eqnarray}
Each function on $\cal M$ can be regarded as a function of (the
function) $y$. Then $f(y) = \sum_k f(y) \, e_k = \sum_k f(q^k) \, e_k$.
Applying $d$ to this expression leads to
\begin{eqnarray}
    df(y) &=& \sum_{k, \epsilon} f(q^k) \, (e_{k-\epsilon} - e_k) \,
            e^\epsilon
          = \sum_{k, \epsilon} \lbrack f(q^{k+\epsilon})-f(q^k)
            \rbrack \, e_k \, e^\epsilon
          = \sum_\epsilon \lbrack f(q^\epsilon y)-f(y) \rbrack \,
            e^\epsilon              \nonumber \\
          &=:& \sum_\epsilon (q^\epsilon -1) \, y \,
               \partial_\epsilon f(y) \; e^\epsilon
\end{eqnarray}
where the `partial derivatives' defined via the last equality are
$q$-derivatives. From (\ref{dy}) and (\ref{tau}) we obtain
\begin{eqnarray}
  e^\epsilon = (q^\epsilon - q^{-\epsilon})^{-1} \, y^{-1} \, dy
               + (1+ q^\epsilon)^{-1} \, (q^{1/2} - q^{-1/2})^{-2}
               \, y^{-2} \, \tau   \; .
\end{eqnarray}
Inserting this into the above expression for $df$ we find
\begin{eqnarray}
    df(y) &=& \sum_\epsilon \partial_\epsilon f(y) \, \lbrack
             {q^\epsilon -1 \over q^\epsilon - q^{-\epsilon}} \, dy
             + {q^\epsilon -1 \over q^\epsilon + 1} \,
             (q^{1/2} - q^{-1/2})^{-2} \, y^{-1} \, \tau
             \rbrack   \nonumber \\
          &=& \bar{\partial}f(y) \, dy + {1 \over 2} \, \Delta f(y)
              \, \tau                         \label{df-dy-tau}
\end{eqnarray}
where in the last step we have introduced the symmetric $q$-derivative
and the $q$-Laplacian,
\begin{eqnarray}
   \bar{\partial} f(y) &:=& {f(q y) - f(q^{-1} y) \over (q-q^{-1})
                            \, y }        \\
   \Delta f(y) &:=& 2 \, (q^{1/2} + q^{-1/2})^{-1} \,
                    (q^{1/2} - q^{-1/2})^{-2} \, y^{-2} \, \lbrack
                    q^{-1/2} \, f(q y)       \nonumber \\
               &  & + q^{1/2} \, f(q^{-1} y)
                    - (q^{1/2} + q^{-1/2}) \, f(y) \rbrack   \; .
\end{eqnarray}
With the help of (\ref{df-dy-tau}), (\ref{dy-y}) and (\ref{tau-y})
one can now calculate, e.g., the commutator $\lbrack dy , f(y) \rbrack
= \lbrack df(y) , y \rbrack$. Furthermore, (\ref{tau-y}) obviously
generalizes to
\begin{eqnarray}
 df(y) = \lbrack (q^{1/2} - q^{-1/2})^{-2} \, y^{-2} \, \tau ,
         f(y) \rbrack   \; .
\end{eqnarray}
                                    \hfill  {\Large $\Box$}

\vskip.3cm
\noindent
Let $({\cal M}, \Omega({\cal M}))$ be a discrete differential manifold
and ${\cal M}'$ a subset of $\cal M$. To ${\cal M}'$ corresponds a
subdiagram of the oriented Hasse diagram for $\Omega({\cal M})$
which then defines a differential calculus $\Omega({\cal M}')$
on ${\cal M}'$. $({\cal M}', \Omega({\cal M}'))$ is a {\em discrete
differential submanifold} of $({\cal M}, \Omega({\cal M}))$.
\vskip.3cm
\noindent
{\bf Example 6.} Fig. 1 shows the digraph of a differential
calculus on a 3-point set. It generates the oriented Hasse diagram
drawn to the right of the digraph. The black points in the digraph
select a 2-point subset. In the depicted oriented Hasse diagram the
corresponding subgraph is emphasized. It is the oriented Hasse diagram
generated by the subdigraph with two points and one arrow.

\hspace*{1.6cm}
\begin{minipage}[t]{3.5cm}
\unitlength2.cm
\begin{picture}(2.,1.5)
\put(0.,1.) {\circle*{0.1}}
\put(1.,0.) {\circle*{0.1}}
\put(-1.,0.) {\circle{0.1}}
\thinlines
\put(-1.,0.) {\vector(1,0){1.9}}
\put(-1.,0.) {\vector(1,1){0.9}}
\thicklines
\put(1.,0.) {\vector(-1,1){0.9}}
\end{picture}
\end{minipage}
\begin{minipage}{4.6cm}
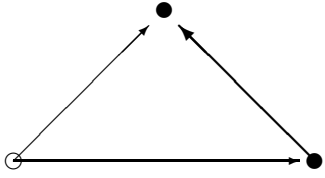
\centerline{\bf Fig. 1}
\vskip.1cm \noindent
\small
A (sub)digraph corresponding to a differential calculus on a
3-point (2-point) set and the corresponding oriented (sub-)
Hasse diagram.
\end{minipage}
\hspace{3.cm}
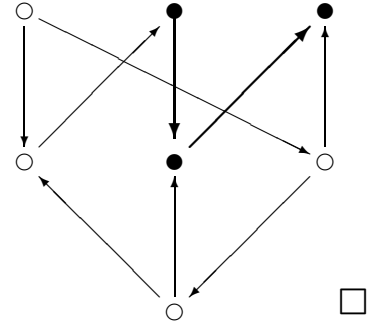
\begin{minipage}{2.6cm}
\unitlength2.cm
\begin{picture}(1.,0.)
\put(-1.,1.) {\circle{0.1}}
\put(0.,1.) {\circle*{0.1}}
\put(1.,1.) {\circle*{0.1}}
\put(-1.,0.) {\circle{0.1}}
\put(0.,0.) {\circle*{0.1}}
\put(1.,0.) {\circle{0.1}}
\put(0.,-1.) {\circle{0.1}}
\thinlines
\put(-1.,0.9) {\vector(0,-1){0.8}}
\put(1.,0.1) {\vector(0,1){0.8}}
\put(-0.9,0.1) {\vector(1,1){0.8}}
\put(-.9,0.95) {\vector(2,-1){1.8}}
\put(0.,-0.9) {\vector(0,1){0.8}}
\put(-0.1,-0.9) {\vector(-1,1){0.8}}
\put(0.9,-0.1) {\vector(-1,-1){0.8}}
\thicklines
\put(0.,1.) {\vector(0,-1){0.85}}
\put(0.1,0.1) {\vector(1,1){0.8}}
\end{picture}
\end{minipage}

                  \hfill  {\Large $\Box$}

\vskip.7cm
\noindent
Let $({\cal M}, \Omega({\cal M}))$ and $({\cal M}', \Omega({\cal M}'))$
be two discrete differential manifolds. From these one can build the
{\em product manifold} $({\cal M} \times {\cal M}',
\Omega({\cal M} \times {\cal M}'))$ where ${\cal M} \times {\cal M}'$
is the cartesian product of the two sets ${\cal M}, {\cal M}'$
and
\begin{eqnarray}
 \Omega({\cal M} \times {\cal M}') := \Omega({\cal M}) \hat{\otimes}
                                      \Omega({\cal M}')
\end{eqnarray}
is the skew tensor product of the two differential algebras (cf
\cite{Kast88}, Appendix A). The product in $\Omega({\cal M} \times
{\cal M}')$ is
\begin{eqnarray}
  (\omega \hat{\otimes} \omega') (\rho \hat{\otimes} \rho')
       = (-1)^{\partial \omega' \cdot \partial \rho} \,
         ( \omega \rho \hat{\otimes} \omega' \rho' )
\end{eqnarray}
and the operator $d$ on $\Omega({\cal M} \times {\cal M}')$ is given by
\begin{eqnarray}
  d (\omega \hat{\otimes} \omega') = (d \omega) \hat{\otimes} \omega'
      + (-1)^{\partial \omega} \, \omega \hat{\otimes} d' \omega' \; .
\end{eqnarray}
Here $\partial \omega$ denotes the grade of the form $\omega$.
The discrete differential manifold $(\Ir^n, \Omega(\Ir^n))$ in example
2 is the $n$-fold product of $(\Ir, \Omega(\Ir))$ from example 3.
Also in example 4 we have an $n$-fold product manifold.
\vskip.3cm
\noindent
The Euler-Poincar{\'e} theorem (see \cite{Croo78}, for example)
suggests the following definition.
\begin{eqnarray}
  \chi (\Omega({\cal M})) := \sum_{r \geq 0} (-1)^r \,
                             \mbox{dim}_\Cx \, \Omega^r({\cal M})
\end{eqnarray}
is the {\em Euler characteristic} of the discrete differential
manifold $({\cal M}, \Omega({\cal M}))$.

\section{Differentiable maps between discrete differential manifolds}
\setcounter{equation}{0}
Let $\phi$ be a map from a discrete differential manifold $({\cal N},
\Omega({\cal N}), d_{\cal N})$ to another one, $({\cal M},
\Omega({\cal M}), d_{\cal M})$.\footnote{The notation $d_{\cal M}$ is
actually a bit misleading. For different differential calculi on $\cal
M$ also the associated operators $d_{\cal M}$ are different.} We define
\begin{eqnarray}
    \phi^\star f := f \circ \phi              \label{phi-star}
\end{eqnarray}
where $f \in {\cal A}_{\cal M}$ and $\circ$ denotes composition of
maps.
\vskip.3cm
\noindent
{\bf Definition.} The map $\phi$ is called {\em differentiable} if
$\phi^\star$ can be consistently extended to a homomorphism of
the corresponding differential algebras, i.e. a linear map
$\Omega({\cal M}) \rightarrow \Omega({\cal N})$ such that
\begin{eqnarray}
  \phi^\star (\omega \omega') &=& \phi^\star (\omega) \,
           \phi^\star (\omega')   \label{star-1}  \\
  \phi^\star d_{\cal M} &=& d_{\cal N} \phi^\star   \; . \label{star-2}
\end{eqnarray}
We refer to $\phi^\star$ as the {\em pull-back} map. Clearly, if
$\phi^\star$ exists, then it is unique.

\vskip.3cm
\noindent
{\bf Lemma.} The composition of differentiable maps is
             differentiable.
\vskip.1cm \noindent
{\bf Proof:} Let $\phi = \rho \circ \gamma$ be the composition of
two differentiable maps ${\cal N} \stackrel{\gamma}{\rightarrow}
{\cal M} \stackrel{\rho}{\rightarrow} {\cal P}$. With
\begin{eqnarray*}
     \phi^\star := \gamma^\star \circ \rho^\star
\end{eqnarray*}
it is easy to verify that (\ref{phi-star}), (\ref{star-1})
and (\ref{star-2}) hold.
                                        \hfill  {\Large $\Box$}

\vskip.3cm
\noindent
{\bf Lemma.} Any map $\phi$ into a set $\cal M$ supplied with the
universal differential calculus $\tilde{\Omega}({\cal M})$ is
differentiable.
\vskip.1cm \noindent
{\bf Proof:} On the algebra ${\cal A}_{\cal M}$ of functions on
$\cal M$ the pull-back $\phi^\star$ is a homomorphism to the algebra
${\cal A}_{\cal N}$ of functions on $\cal N$. A general result
(see \cite{Kast88}, Proposition 1.8) then tells us that there is
a unique extension to a homomorphism $\tilde{\Omega}({\cal M})
\rightarrow \Omega({\cal N})$. More concretely, one
defines
\begin{eqnarray*}
       \phi^\star (f_0 \, d_{\cal M} f_1 \cdots d_{\cal M} f_r)
    := (\phi^\star f_0) \, (d_{\cal N} \phi^\star f_1) \cdots
       (d_{\cal N} \phi^\star f_r)
\end{eqnarray*}
($\forall r \in \Nl, \, f_s \in {\cal A}_{\cal M}$). By linearity
and the Leibniz rule $(df) \, h = d(fh) - f \, dh$ for $f,h \in
{\cal A}_{\cal M}$, $\phi^\star$ is then defined on arbitrary
forms. It can now be shown that (\ref{star-1}) and (\ref{star-2})
are satisfied for arbitrary forms (see \cite{Kast88}, Proposition
1.8, for details).
                                             \hfill  {\Large $\Box$}

\vskip.3cm
\noindent
We recall that a differential algebra $\Omega({\cal M})$ is the image
of a homomorphism $\pi \, : \, \tilde{\Omega}({\cal M}) \rightarrow
\Omega({\cal M})$ and therefore the quotient of the universal
differential algebra $\tilde{\Omega}({\cal M})$ by some differential
ideal ${\cal I}_{\cal M}$. A useful characterization of
differentiability is now given in the next Lemma.
\vskip.3cm
\noindent
{\bf Lemma.} A map $\phi \, : \, {\cal N} \rightarrow {\cal M}$ is
differentiable with respect to differential calculi $\Omega({\cal N})$
and $\Omega({\cal M}) = \tilde{\Omega}({\cal M})/{\cal I}_{\cal M}$ iff
\begin{eqnarray}
   \tilde{\phi}^\ast ({\cal I}_{\cal M}) = 0       \label{ideal}
\end{eqnarray}
where $\tilde{\phi}^\star$ is the homomorphism $\tilde{\Omega}({\cal M})
\rightarrow \Omega({\cal N})$ which exists according to the
previous Lemma.
\vskip.1cm \noindent
{\bf Proof:}
``$\Rightarrow$'': If $\phi$ is differentiable so that $\phi^\star$
extends to a homomorphism $\Omega({\cal M}) \rightarrow
\Omega({\cal N})$, then $\tilde{\phi}^\star := \phi^\star \circ \pi$
extends $\phi^\star \, : \, {\cal A}_{\cal M} \rightarrow
{\cal A}_{\cal N}$ to a homomorphism $\tilde{\Omega}({\cal M})
\rightarrow \Omega({\cal N})$. It must then coincide with the map
$\tilde{\Omega}({\cal M}) \rightarrow \Omega({\cal N})$ which always
exists according to the last Lemma. By definition of $\pi$ we have
$\tilde{\phi}^\ast ({\cal I}_{\cal M}) = 0$.    \\
``$\Leftarrow$'': According to the previous Lemma the homomorphism
$\phi^\star \, : \, {\cal A}_{\cal M} \rightarrow {\cal A}_{\cal N}$
lifts to a homomorphism $\tilde{\phi}^\star \, : \,
\tilde{\Omega}({\cal M}) \rightarrow \Omega({\cal N})$ of
differential algebras. Using a general result in algebra (see
\cite{Hung74}, for example), this map induces a homomorphism
$\phi^\star \, : \, \Omega({\cal M}) \rightarrow \Omega({\cal N})$
of differential algebras if (\ref{ideal}) holds.
                                      \hfill  {\Large $\Box$}

\vskip.3cm
\noindent
Let $\Omega({\cal M})$ be a reduction of $\tilde{\Omega}({\cal M})$
such that $e_{ab} = 0$ (respectively $\pi(e_{ab}) = 0$ refering to
$e_{ab} \in \tilde{\Omega}({\cal M})$). For differentiable $\phi$ we
then have
\begin{eqnarray}                    \label{tildephi}
  0 = \tilde{\phi}^\star (e_{ab})
    = (\phi^\star e_a) \, d_{\cal N} (\phi^\star e_b)
    = \sum_{i \in \phi^{-1}(a)} e_i \; d_{\cal N}
      \sum_{j \in \phi^{-1}(b)} e_j
    = \sum_{i \in \phi^{-1}(a) \atop j \in \phi^{-1}(b)} e_{ij}
\end{eqnarray}
where we have used (\ref{star-1}), (\ref{star-2}) and
\begin{eqnarray}                      \label{ecircphi}
       e_a \circ \phi = \sum_{i \in \phi^{-1}(a)} e_i
\end{eqnarray}
($i,j, \ldots$ denote elements of $\cal N$).
For a given map $\phi$ we may regard (\ref{tildephi}) as a constraint on
the differential calculi on $\cal N$ and $\cal M$ which are needed to
render a map $\phi$ differentiable. It has a simple interpretation in
terms of the digraphs associated with the differential calculi. If two
points of $\cal N$ connected by an arrow are mapped into two different
points, then -- in order for $\phi$ to be differentiable -- there must
be an arrow between the image points with the same orientation. If we
fix differential calculi on $\cal N$ and $\cal M$, respectively, the
differentiability restricts the allowed class of maps, of course.
Corresponding examples are given in the following section.
\vskip.3cm

More generally we may have reductions on the level of $r$-forms.
Differentiability conditions for $\phi$ are then obtained by using the
general formula
\begin{eqnarray}
      \tilde{\phi}^\star (e_{a_1 \ldots a_{r+1}})
    = \sum_{i_1 \in \phi^{-1}(a_1) \atop {\vdots \atop i_{r+1}
      \in \phi^{-1}(a_{r+1})}} e_{i_1 \ldots i_{r+1}}  \; .
\end{eqnarray}

\section{Diffeomorphisms of a discrete differential manifold}
\setcounter{equation}{0}
Let ${\cal N} = {\cal M}$ and $\phi$ a bijection. If $\phi$ is
differentiable as a map $({\cal M}, \Omega'({\cal M})) \rightarrow
({\cal M}, \Omega({\cal M}))$, then $\phi^{-1}$ is not differentiable,
in general. If we want both, $\phi$ and $\phi^{-1}$, differentiable
then we must have
\begin{eqnarray}           \label{diffeo1}
     e_{\phi(a) \phi(b)} \neq 0 \quad \Leftrightarrow \quad
     e_{ab} \neq 0
\end{eqnarray}
and corresponding conditions in case of higher order reductions. This
is only possible if $\Omega'({\cal M}) = \Omega({\cal M})$.
We call a bijection $\phi$ a {\em diffeomorphism} of a discrete
differential manifold $({\cal M}, \Omega({\cal M}))$ if $\phi$ and
$\phi^{-1}$ are differentiable with respect to $\Omega({\cal M})$.
\vskip.3cm
\noindent
{\bf Lemma.} Let $\cal M$ be a finite set and $\phi$ a bijection which
is differentiable with respect to a first order reduction
$\Omega({\cal M})$ of the universal differential calculus on $\cal M$.
Then $\phi^{-1}$ is also differentiable (and therefore a diffeomorphism).
\vskip.1cm \noindent
{\bf Proof:} The differentiability of $\phi$ implies that
$e_{ij} \neq 0 \; \Rightarrow \; e_{\phi(i) \phi(j)} \neq 0$ so
that an arrow in the digraph of the differential calculus which points
from $i$ to $j$ is mapped into an arrow from $\phi(i)$ to $\phi(j)$.
But also the inverse implication $e_{\phi(i) \phi(j)} \neq 0 \;
\Rightarrow \; e_{ij} \neq 0$ holds since otherwise the map would
`create' an arrow and thus change the differential calculus. Hence
$e_{ij} = 0 \; \Leftrightarrow \; e_{\phi(i) \phi(j)} = 0$.
Now the statement follows using the last Lemma of the previous
section.
                              \hfill  {\Large $\Box$}

\vskip.3cm
\noindent
The statement in the Lemma is not true for infinite sets, in general.
\vskip.3cm
\noindent
The {\em adjacency matrix} $A = (A_{ij})$ of a digraph $G$ is defined
by
\begin{eqnarray}
  A_{ij} = \left \lbrace        \begin{array}{ll}
   1 \quad & \mbox{if there is an arrow from } i \mbox{ to } j \\
   0       & \mbox{otherwise}   \end{array}
           \right.
\end{eqnarray}
{}From graph theory we recall the following characterization of an
automorphism of a digraph $G$ \cite{Chao65}. A bijection $\phi$ is
an element of the automorphism group $\mbox{Aut}(G)$ of $G$ iff the
associated matrix
\begin{eqnarray}
               (M_\phi)_{ij} = \delta_{\phi^{-1}(i) \, j}
\end{eqnarray}
commutes with the adjacency matrix of $G$, i.e.
\begin{eqnarray}
        \lbrack M_\phi , A \rbrack = 0     \; .
\end{eqnarray}

\vskip.3cm
\noindent
{\bf Proposition.} Let $\Omega({\cal M})$ be a first order reduction
of the universal differential calculus on $\cal M$ with adjacency
matrix $A$. A bijection $\phi$ is a diffeomorphism iff it is an
automorphism of the corresponding digraph.
\vskip.1cm \noindent
{\bf Proof:}  Since
\begin{eqnarray*}
   (M_\phi A)_{ij} = A_{\phi^{-1}(i) j} \quad , \quad
   (A \, M_\phi)_{ij} = A_{i \phi(j)}
\end{eqnarray*}
the condition $\lbrack M_\phi , A \rbrack = 0$ is equivalent to
\begin{eqnarray*}
              A_{ij} = A_{\phi(i) \phi(j)}
\end{eqnarray*}
respectively,
\begin{eqnarray*}
   A_{ij} = 0 \quad \Leftrightarrow \quad
            A_{\phi(i) \phi(j)} = 0 \; .
\end{eqnarray*}
But this in turn is equivalent to
\begin{eqnarray*}
   e_{ij} \in {\cal I} \quad \Leftrightarrow \quad
            e_{\phi(i) \phi(j)} \in {\cal I}
\end{eqnarray*}
if we express $\Omega({\cal M}) = \tilde{\Omega}({\cal M})/{\cal I}$.
The last statement is equivalent to the differentiability of $\phi$
and $\phi^{-1}$.
                              \hfill  {\Large $\Box$}

\vskip.3cm
\noindent
For a digraph with a finite number $N$ of vertices, the automorphism
group $\mbox{Aut}(G)$ is a subgroup of the symmetric group $S_N$ (the
group of permutations).
For the first graph in Fig. 2  the automorphism group consists of
the identity only. The second graph obviously has a (discrete)
rotational symmetry. In this case we have $\mbox{Aut}(G) \cong
\Ir_3$.

\hspace*{1.6cm}
\begin{minipage}[t]{3.cm}
\unitlength2.cm
\begin{picture}(2.,1.5)
\thicklines
\put(0.,1.) {\circle*{0.1}}
\put(1.,0.) {\circle*{0.1}}
\put(-1.,0.) {\circle*{0.1}}
\put(-1.,0.) {\vector(1,0){1.9}}
\put(1.,0.) {\vector(-1,1){0.9}}
\put(-1.,0.) {\vector(1,1){0.9}}
\end{picture}
\end{minipage}
\begin{minipage}{5cm}
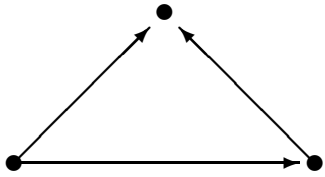
\centerline{\bf Fig. 2}
\vskip.1cm \noindent
\small
The digraphs corresponding to two different differential
calculi on a three point set.
\vspace{2cm}
\end{minipage}
\hspace{2.4cm}
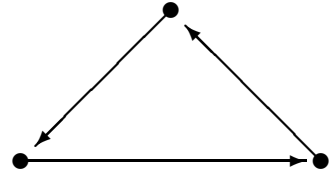
\begin{minipage}{3.cm}
\unitlength2.cm
\begin{picture}(2.,0.)
\thicklines
\put(0.,1.) {\circle*{0.1}}
\put(1.,0.) {\circle*{0.1}}
\put(-1.,0.) {\circle*{0.1}}
\put(-1.,0.) {\vector(1,0){1.9}}
\put(1.,0.) {\vector(-1,1){0.9}}
\put(0.,1.) {\vector(-1,-1){0.9}}
\end{picture}
\end{minipage}

\section{Differentiable curves in discrete differential manifolds}
\setcounter{equation}{0}
A 1-dimensional discrete differential manifold has been described
in example 3 in section III. Its differential operator will be
denoted by $d$ in the following. A {\em differentiable curve} in a
discrete differential manifold $\cal M$ should then be a
differentiable map $\gamma$ from $\Ir$ (with the differential calculus
of example 3 in section III) to $\cal M$ (with its differential
calculus). Instead of $\Ir$ we may take as well ${\bf N} = \lbrace 0,
\ldots N-1 \rbrace$ (with the induced differential calculus).
\vskip.3cm
\noindent
{\bf Example 1.} Let ${\cal M} = \Ir^n$ with the differential calculus
of example 2 in section III, i.e. $\tilde{\Omega}({\cal M})/{\cal I}
_{\cal M}$ where the ideal ${\cal I} _{\cal M}$ is generated by
\begin{eqnarray}
  \lbrack d_{\cal M} x^\mu , x^\nu \rbrack - \delta^{\mu \nu} \;
     d_{\cal M} x^\nu   \qquad (\mu, \nu = 1, \ldots, n)  \; .
\end{eqnarray}
According to the last Lemma in section IV, $\gamma \, :
\, \Ir \rightarrow {\cal M}$ is differentiable if $\tilde{\gamma}^\star$
maps the last expression to the zero in $\Omega(\Ir)$ so that
\begin{eqnarray}
 0 = \tilde{\gamma}^\star (\lbrack d_{\cal M} x^\mu , x^\nu \rbrack
     - \delta^{\mu \nu} \; d_{\cal M} x^\nu)
   = \lbrack d x^\mu(t) , x^\nu(t) \rbrack - \delta^{\mu \nu}
     \; d x^\nu(t)
\end{eqnarray}
where $x^\mu(t) := x^\mu \circ \gamma (t)$. Now (\ref{df-h})
leads to the differentiability condition
\begin{eqnarray}
   \dot{x}^\mu(t) \, \lbrack \dot{x}^\nu(t) - \delta^{\mu \nu} \rbrack
   = 0    \; .
\end{eqnarray}
If we regard $\gamma$ as a curve describing the motion of a particle
on the lattice $\Ir^n$, the last condition restricts the motion
such that the particle can either rest at a given site or hop to a
neighboring site. Furthermore, only a motion to a site with increasing
values of $x^\mu$ is allowed. This apparently unplausible restriction
is absent in our next example. It reminds us, however, of right (and
left) movers in 2-dimensional chiral field theories. If we had taken the
differential calculus on $\Ir^n$ with the opposite orientation of arrows
in the corresponding digraph, then only motion to a site with decreasing
values of $x^\mu$ would be allowed. There are, of course, $2n$ ways of
choosing the direction of arrows along the $n$ axes of $\Ir^n$ and
thus $2^n$ `chiral sectors' in the lattice. In order to reach (in
principle) all lattice sites, we would need $2^n$ particles, each moving
in a separate chiral sector of the lattice (directed by a separate
differential calculus). This reminds us of the
fermion doubling problem in lattice theories.\footnote{We refer to
\cite{Roth92} for a discussion of this problem. See also
\cite{tHoo88} for the relation between left and right movers on a
one-dimensional lattice and the one-dimensional free massless fermion
gas.}
The problem is solved if we take the symmetric lattice (see next
example) on which a single particle can move in each lattice direction.
                                        \hfill  {\Large $\Box$}

\vskip.3cm
\noindent
{\bf Example 2.} Again we choose ${\cal M} = \Ir^n$, but now
with the differential calculus determined by (\ref{xdx-symlatt}),
i.e. we are dealing with the `symmetric lattice' calculus of
\cite{DMH94}. Differentiability of $\gamma \, : \,  \Ir \rightarrow
{\cal M}$ requires
\begin{eqnarray}
 \dot{x}^\mu(t) \, \lbrack d t , x^\nu(t) \rbrack
        = \delta^{\mu \nu} \, w^\nu(t) \, d t    \qquad
 w^\mu(t) \, \lbrack d t , x^\nu(t) \rbrack
 = \delta^{\mu \nu} \, \dot{x}^\nu(t) \, d t
\end{eqnarray}
where $w^\mu(t)$ is given by $\gamma^\star \tau^\mu = w^\mu(t) \, d t$.
{}From these equations one derives
\begin{eqnarray}
  \dot{x}^\mu \, \dot{x}^\nu = \delta^{\mu \nu} \, w^\nu  \qquad
  w^\mu \, \dot{x}^\nu = \delta^{\mu \nu} \, \dot{x}^\nu  \; .
\end{eqnarray}
The first equation implies $\dot{x}^\mu \, \dot{x}^\nu = 0$
for $\mu \neq \nu$ so that at most one $\dot{x}^\kappa$ (with fixed
$\kappa$) can be different from zero (at a given value of $t$).
The above equations then reduce to $\dot{x}^\kappa = \pm 1$. Hence,
the `particle' is allowed to jump to an arbitrary neighboring site on
the lattice. The remaining solution of the above conditions, namely
$\dot{x}^\mu = 0$ for all $\mu$, allows the particle to remain at
a site.                                        \hfill  {\Large $\Box$}

\vskip.3cm
\noindent
{\bf Example 3.} Now we choose ${\cal M} = \Ir_N$ with the
differential calculus of example 5 in section III. If $\gamma \, :
\, \Ir \rightarrow \Ir_N$ is a differentiable curve, then the
pull-back of (\ref{dy-y}) and (\ref{tau-y}) yields
\begin{eqnarray}
  \dot{y}(t)^2 = (q^{1/2} - q^{-1/2})^2 \, y(t) \, \dot{y}(t) + w(t)
  \qquad \quad
  w(t) \, \dot{y}(t) = (q^{1/2} - q^{-1/2})^2 \, y(t)^2 \, \dot{y}(t)
\end{eqnarray}
where $dy(t) =: \dot{y}(t) \, dt$ and $\gamma^\star \tau =: w \, dt$.
If $\dot{y}$ vanishes at a certain `time', then $w = 0$ at that
time. If $\dot{y} \neq 0$ (at some time), the above equations imply
$w = (q^{1/2} - q^{-1/2})^2 \, y(t)^2$ and restrict $\dot{y}$ to the
values $(q^\epsilon-1) \, y$ where $\epsilon = \pm 1$. All these
conditions can now be summarized in the formula
\begin{eqnarray}
  \dot{y} \, \lbrack \dot{y} - (q-1) \, y \rbrack \,
             \lbrack \dot{y} - (q^{-1}-1) \, y \rbrack = 0  \; .
\end{eqnarray}
Using $\dot{y}(t) = y(t+1) - y(t)$, this means
\begin{eqnarray}
  y(t+1) = y(t) \quad \mbox{or} \quad q \, y(t)
                \quad \mbox{or} \quad q^{-1} \, y(t)   \; .
\end{eqnarray}
The particle thus either remains at a site or moves one step on the
periodic lattice (which the coordinate $y$ describes as a
$q$-lattice).                                \hfill  {\Large $\Box$}

\vskip.3cm
\noindent
In all these examples the particle can only move one lattice spacing
in at least one time step. This means that there is a maximal
velocity which we may identify with the vacuum velocity of light.
It should be noticed, however, that such an interpretation presumes
that there is a time and a space metric. A natural choice is indeed
suggested by our description of discrete time as $\Ir$ (or $\bf N$)
and discrete space as (a subset of) $\Ir^n$. But these are extra
structures which we still have to introduce on discrete sets and to
discuss in more generality.

\vskip.3cm
\noindent
{\bf Example 4.} Let $\gamma$ be a curve in $\Ir^n$ subject to the
equation of motion
\begin{eqnarray}
              \partial_- \partial_+ x^\mu(t) = 0  \; .
\end{eqnarray}
With (\ref{partial+-}) this becomes
\begin{eqnarray}
                x(t+1) - 2 \, x(t) + x(t-1) = 0
\end{eqnarray}
which implies
\begin{eqnarray}
                x(t+1) - x(t) = x(t) - x(t-1) \; .
\end{eqnarray}
Hence, in each time step the distance traversed on $\Ir^n$
(in each of the $n$ canonical lattice directions) is constant. This
corresponds to our intuitive conception of a
free motion on a lattice. Imposing differentiability of $\gamma$
with respect to some choice of differential calculus on $\Ir^n$
now further restricts these motions. A motion on such a discrete
differentiable manifold is then only allowed if the associated
digraph has an arrow between every two adjacent points along the
traversed path which points in the direction of the motion. A particular
consequence is the existence of a maximal velocity (as already pointed
out). In case of the oriented lattice calculus (example 1) there are
only two free differentiable motions. Either the particle remains
forever at one site or it moves steadily (in one time step) to the
neighboring site in one of the canonical lattice directions. For
motion on the one-dimensional oriented lattice this is illustrated
in Fig. 3.

\begin{minipage}{7cm}
\unitlength1.cm
\begin{picture}(7,6)
\thinlines
\multiput(0,0)(1,0){7}{\circle*{0.15}}
\multiput(0,1)(1,0){7}{\circle*{0.15}}
\multiput(0,2)(1,0){7}{\circle*{0.15}}
\multiput(0,3)(1,0){7}{\circle*{0.15}}
\multiput(0,4)(1,0){7}{\circle*{0.15}}
\multiput(0,5)(1,0){7}{\circle*{0.15}}
\multiput(0,0)(1,0){6}{\vector(1,0){0.9}}
\multiput(0,1)(1,0){6}{\vector(1,0){0.9}}
\multiput(0,2)(1,0){6}{\vector(1,0){0.9}}
\multiput(0,3)(1,0){6}{\vector(1,0){0.9}}
\multiput(0,4)(1,0){6}{\vector(1,0){0.9}}
\multiput(0,5)(1,0){6}{\vector(1,0){0.9}}
\thicklines
\multiput(1,0)(0,1){5}{\vector(0,1){0.9}}
\multiput(2,0)(1,1){4}{\vector(1,1){0.9}}
\end{picture}
\end{minipage}
\hspace{2.5cm}
\begin{minipage}[t]{5cm}
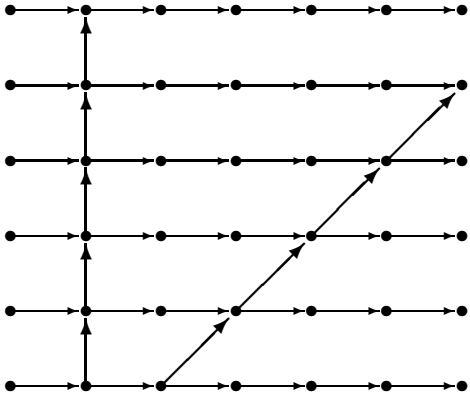
\centerline{\bf Fig. 3}
\vskip.1cm \noindent
\small
Free differentiable motions on the one-dimensional oriented lattice
in a space-time picture.
\end{minipage}

                                             \hfill  {\Large $\Box$}

\vskip.3cm
Again, let $\gamma$ be a differentiable curve in a discrete
differential manifold $({\cal M}, \Omega({\cal M}))$. Acting with
$\tilde{\gamma}^\star$ (cf (\ref{tildephi})) on $e_a \, e_b =
\delta_{ab} \, e_b$ and $\sum_{a} e_a = \idty$ we obtain
\begin{eqnarray}            \label{e(t)-rels}
  e_a(t) \, e_b(t) = \delta_{ab} \, e_b(t)
            \qquad
  \sum_{a} e_a(t) = \idty (t) = 1
             \qquad (a,b \in {\cal M})
\end{eqnarray}
where $e_a(t) := e_a(\gamma(t))$.
If an arrow from $a$ to a different point $b$ is missing in the
digraph associated with a differential calculus on $\cal M$, then
$\tilde{\gamma}^\star (e_{ab}) = 0$ (cf (\ref{tildephi})) and the
homomorphism property of $\tilde{\gamma}$ leads to
\begin{eqnarray}
    0 = e_a(t) \, \dot{e}_b(t)
      =  e_a(t) \, \lbrack e_b(t+1) - e_b(t) \rbrack
      =  e_a(t) \, e_b(t+1)           \label{tt+1}
\end{eqnarray}
where we used (\ref{e(t)-rels}).

\vskip.3cm
As a {\em Lagrangian} for a dynamical system on a discrete differential
manifold we may regard a function
\begin{eqnarray}
   L\lbrack \gamma \rbrack (t) =  L'(e_a(t), \dot{e}_b(t),
   \ddot{e}_c(t), \ldots )
\end{eqnarray}
with suitably defined second and higher order derivatives
of $e_a(t)$. If there are no higher than first order derivatives,
it can be rewritten as
\begin{eqnarray}
   L\lbrack \gamma \rbrack (t) =  L(e_a(t), e_b(t+1))
   = \sum_{a,b} L_{ab} \, e_a(t) \, e_b(t+1)    \; .
\end{eqnarray}
The last expression gives the most general form of a (first order)
Lagrangian. Note that it also allows terms linear in $e_a(t)$ (or
$e_b(t+1)$) since we have the relation $\sum_a e_a(t) = \idty$.

\section{The space of differentiable curves}
\setcounter{equation}{0}
We recall that a digraph $({\cal M}_0, {\cal M}_1)$ consists of a set
${\cal M} := {\cal M}_0$ of vertices and a set ${\cal M}_1$ of arrows.
In the present context only a subclass of digraphs is considered.
No multiple arrows are allowed between the same pair of vertices. Also
no loops are allowed which forbids arrows originating and ending at the
same vertex. We already know that any digraph defines a differential
calculus on $\cal M$ and vice versa. From the set ${\cal M}_1$ of arrows
one can construct the set ${\cal M}_r$ of paths $(a_0 \to a_1 \to \cdots
\to a_r)$ of length $r$ with $(a_{k-1} \to a_k) \in {\cal M}_1$. The
vertices are paths of length $0$, arrows (elements of ${\cal M}_1$) are
paths of length $1$. For $a,b \in {\cal M}_0$, let
$\ell_{ab}$ denote the minimal length of paths from $a$ to $b$.
In terms of the adjacency matrix,
\begin{eqnarray}
    \ell_{ab} = \mbox{min} \lbrace \ell \, \mid \, (A^\ell)_{ab} \neq
                0 \rbrace
\end{eqnarray}
since $(A^\ell)_{ab}$ is the number of paths of length $\ell$ from
$a$ to $b$. If there is no path from $a$ to $b$, we set $\ell_{ab} =
\infty$.\footnote{Note that $\ell_{ab}$ is not a distance function on
$\cal M$ (as considered, for example, in \cite{Anto94,Alva88} in a
context related to our work) since it is directed, i.e. $\ell_{ab}
\neq \ell_{ba}$, in general.}
\vskip.3cm

The formulae (\ref{e(t)-rels}) and (\ref{tt+1}) motivate the
following algebraic construction. Let $\tilde{\cal A}({\bf N},{\cal M})$
be the commutative and associative algebra generated by elements
$e_a(i), \, a \in {\cal M}, \, i \in {\bf N} = \lbrace 0, \ldots N-1
\rbrace$ subject to the relations
\begin{eqnarray}            \label{e(i)-rels}
  e_a(i) \, e_b(i) = \delta_{ab} \, e_b(i)
            \qquad
  \sum_{a} e_a(i) = 1
             \qquad (a,b \in {\cal M}, \, i \in {\bf N})   \; .
\end{eqnarray}
As we will see in the following, $\tilde{\cal A}({\bf N},{\cal M})$
may be regarded as the algebra of functions on the space of curves
from $\bf N$ to ${\cal M}$ with the universal differential calculus
corresponding to the complete digraph (which has exactly two
antiparallel arrows between all pairs of vertices representing elements
of $\cal M$). More generally, we will associate an algebra
${\cal A}({\bf N},{\cal M})$ with any digraph (with set of vertices
$\cal M$).
\vskip.3cm

Let $\cal J$ denote the two-sided ideal in $\tilde{\cal A}({\bf N}
,{\cal M})$ generated by those products $e_a(i) e_b(j)$ for which
$0 \leq j-i < \ell_{ab}$ for a given digraph. We define
\begin{eqnarray}
  {\cal A}({\bf N},{\cal M}) := \tilde{\cal A}({\bf N},{\cal M})
                                / {\cal J}   \; .
\end{eqnarray}
In case of the complete digraph we have $\ell_{ab} \in \lbrace 0,1
\rbrace$ ($\forall a,b \in {\cal M}, \, a \neq b$) and thus
${\cal J} = 0$.
\vskip.3cm

{}From the last section we infer that any differentiable curve
$\gamma \, : \, {\bf N} \rightarrow {\cal M}$ defines an irreducible
representation $\rho_\gamma$ of ${\cal A}({\bf N},{\cal M})$ via
\begin{eqnarray}                          \label{repr}
          \rho_\gamma(e_a(i)) = e_a(\gamma(i))  \; .
\end{eqnarray}
Conversely, any irreducible representation of ${\cal A}({\bf N},
{\cal M})$ is a differentiable curve. Indeed, since the algebra is
commutative, all irreducible representations are one-dimensional.
(\ref{e(i)-rels}) then implies $\rho(e_a(i)) \in \lbrace 0,1 \rbrace$
and that for each $i$ there is precisely one $a \in {\cal M}$ for
which $\rho(e_a(i)) = 1$. Hence we have a curve $\gamma \, : \, {\bf N}
\rightarrow {\cal M}$. The relations defining the ideal $\cal J$ now
restrict $\gamma$ to be differentiable.
\vskip.3cm

Let $\Gamma := \lbrace \gamma \, : \, {\bf N} \rightarrow {\cal M}
\mid \gamma \mbox{ differentiable with respect to } \Omega({\cal M})
\rbrace$ be the space of differentiable curves (with respect to some
differential calculus $\Omega({\cal M})$). This is again a discrete
set to which the formalism of section II applies. The algebra
${\cal A}(\Gamma)$ of $\Cx$-valued functions on $\Gamma$ is then
generated by elements $e_\gamma$ such that
\begin{eqnarray}
   e_\alpha(\beta) = \delta_{\alpha \beta} \; , \qquad
   e_\alpha \, e_\beta = \delta_{\alpha \beta} \, e_\beta \; , \qquad
   \sum_{\gamma \in \Gamma} e_\gamma = \idty_\Gamma  \; .
\end{eqnarray}
Via (\ref{repr}) we may regard $e_a(i)$ as a function on $\Gamma$,
\begin{eqnarray}                \label{e_a(i)}
   e_a(i) = \sum_\gamma e_a (\gamma (i)) \, e_\gamma =
            \sum_\gamma \delta_{a \gamma(i)} \, e_\gamma  \; .
\end{eqnarray}
Now
\begin{eqnarray}
   e_\gamma = \prod_{i \in {\bf N}} e_{\gamma(i)}(i)
\end{eqnarray}
shows that ${\cal A}({\bf N},{\cal M}) = {\cal A}(\Gamma)$.
\vskip.3cm

A (first order) {\em action} is a function on $\Gamma$. It can
always be written in the form
\begin{eqnarray}
      S = \sum_i \sum_{a,b} L_{ab}(i) \; e_a(i) \, e_b(i+1)
\end{eqnarray}
with real (or complex) coefficients $L_{ab}(i)$. `Classical motions'
should correspond to (local) extrema of the action. To find a
local extremum a formalism of variations should be helpful. The
latter may be realized as a differential calculus on the space of
curves.

\subsection{Differential calculus on the space of curves}
Let $(\tilde{\Omega}(\Gamma), \tilde{d})$ be the universal
differential calculus on $\Gamma$. For $a \neq b$ we define
\begin{eqnarray}
        e_{ab}(i) := e_a(i) \, \tilde{d} e_b(i)
\end{eqnarray}
and $e_{aa}(i) := 0$. The differential calculus $\Omega({\cal M})
= \tilde{\Omega}({\cal M})/{\cal I}_{\cal M}$ which enters the
definition of $\Gamma$ induces a reduction of $\tilde{\Omega}(\Gamma)$
in the following way. Let ${\cal I}_\Gamma$ be the two-sided
differential ideal of $\tilde{\Omega}(\Gamma)$ generated by those
$e_{ab}(i)$ for which $e_{ab} \in {\cal I}_{\cal M}$. We define
\begin{eqnarray}
  \Omega({\Gamma}) := \tilde{\Omega}(\Gamma)/{\cal I}_\Gamma \; .
\end{eqnarray}
In the associated digraph there is an arrow from $\alpha$ to
$\beta$ (regarded as vertices) iff for all $i \in {\bf N}$ there
is an arrow from $\alpha(i)$ to $\beta(i)$ in the digraph
corresponding to $\Omega({\cal M})$.
Whereas the basic 1-forms $e_{\alpha \beta}$ constitute a basis of
$\Omega^1(\Gamma)$ over $\Cx$, this is not so for the set of 1-forms
$e_{ab}(i)$. The latter is in general not even a basis of
$\Omega^1(\Gamma)$ as a left ${\cal A}(\Gamma)$-module. There are
not enough commutation relations between $e_a(i)$ and $de_b(j)$ for
$i \neq j$.
\vskip.3cm

If there is no arrow from $a$ to $c \neq a$ in the digraph for
$\Omega({\cal M})$, but a path from $a$ to $c$ via one further
vertex $b$, then $e_a(i) \, e_c(i+1) = 0$ (cf the definition of
${\cal A}({\bf N}, {\cal M})$). Acting with $d$ on this equation
one finds
\begin{eqnarray}
  \sum_b \lbrack e_a(i) \, e_{bc}(i+1) - e_{ab}(i) \, e_c(i+1)
              \rbrack = 0  \; .
\end{eqnarray}
Here we have used
\begin{eqnarray}                 \label{de_a(i)}
  d e_a(i) = \sum_{b \in {\cal M}} \lbrack e_{ba}(i) - e_{ab}(i)
             \rbrack
\end{eqnarray}
which follows from (\ref{e_a(i)}) and the general formula
(\ref{de...}).
\vskip.3cm

The problem to determine a local minimum of an action $S$ can now be
formulated as follows. One has to find a curve $\gamma \in \Gamma$
such that
\begin{eqnarray}             \label{localmin}
  dS(\gamma, \alpha) \geq 0  \qquad  \mbox{and} \qquad
  dS(\alpha, \gamma) \leq 0  \qquad \forall \alpha \in \Gamma \; .
\end{eqnarray}
Here we make use of the representation $e_{ab}(i) = e_a(i) \otimes
e_b(i)$ for $e_{ab} \neq 0$. Note that $dS$ can only be nonvanishing
on pairs of neighboring curves.
\vskip.3cm
\noindent
{\bf Remark.} Let us call two curves $\alpha, \beta$ neighbors if
for all $i \in {\bf N}$ either $\alpha(i) = \beta(i)$ or
there is an arrow between $\alpha(i)$ and $\beta(i)$ in the
digraph for $\Omega({\cal M})$. Then, for $\alpha \in \Gamma$
there are, in general, neighboring curves $\beta$ which are
not in $\Gamma$ (i.e., not differentiable). A corresponding extended
space of curves could be of relevance for a calculus of variations
which should determine discrete dynamics from an action.
                            \hfill {\Large $\Box$}

\subsection{A simple example}
Let ${\cal M} = \Ir$ with the oriented lattice calculus (example 3 in
section III). In this case we have
\begin{eqnarray}
                e_a(i) \, e_{a+2}(i+1) = 0
\end{eqnarray}
which implies
\begin{eqnarray}
   e_{a, a+1}(i) \, e_{a+2}(i+1) = e_a(i) \, e_{a+1,a+2}(i+1)
\end{eqnarray}
and
\begin{eqnarray}       \label{de_e_ex}
  d(e_a(i) \, e_{a+1}(i+1)) = e_{a, a+1}(i+1) - e_{a, a+1}(i)  \; .
\end{eqnarray}
In the case under consideration, the most general first order
action takes the form
\begin{eqnarray}
  S = \sum_{a \in \Ir} \left( \sum_{i=0}^{N-2} K_a(i) \, e_a(i) \,
  e_{a+1}(i+1) - \sum_{i=0}^{N-1} V_a(i) \, e_a(i) \right)  \; .
\end{eqnarray}
With the help of (\ref{de_e_ex}) we can calculate its differential,
\begin{eqnarray}
  dS &=& \sum_{a \in \Ir} \left( \sum_{i=0}^{N-2} K_a(i) \,
       \lbrack e_{a,a+1}(i+1) - e_{a,a+1}(i) \rbrack
       - \sum_{i=0}^{N-1} \lbrack V_{a+1}(i) - V_a(i) \rbrack
       \, e_{a,a+1}(i) \right) \nonumber \\
     &=& \sum_{a \in \Ir} \big( - \sum_{i=1}^{N-2} \lbrack K_a(i)
       - K_a(i-1) + V_{a+1}(i) - V_a(i) \rbrack \, e_{a,a+1}(i)
                         \nonumber \\
     & & - \lbrack K_a(0) + V_{a+1}(0) - V_a(0) \rbrack \, e_{a,a+1}(0)
                         \nonumber \\
     & & + \lbrack K_a(N-2) - V_{a+1}(N-1) + V_a(N-1) \rbrack
         \, e_{a,a+1}(N-1) \, \big)   \nonumber  \\
     &=:& \sum_{a \in \Ir} \sum_{i=0}^{N-1} S_a(i) \, e_{a,a+1}(i)
                               \; .
\end{eqnarray}
The inequalities (\ref{localmin}) now read
\begin{eqnarray}
 \sum_{i=0}^{N-1} S_{\gamma(i)}(i) \; \delta_{\gamma(i)+1,\alpha(i)}
    &\geq& 0     \\
 \sum_{i=0}^{N-1} S_{\gamma(i)-1}(i) \; \delta_{\gamma(i)-1,\alpha(i)}
    &\leq& 0
\end{eqnarray}
for all curves $\alpha \in \Gamma$ which are neighbors of $\gamma$.
In the case under consideration there are not enough neighbors in
$\Gamma$ so that we could convert the sums into `local'
inequalities in the sense that they involve at most two time steps.
\vskip.3cm

Let us specify the action by choosing $V_a(i) = 0$ and $K_a(i) = 1$
($\forall a \in \Ir, \, i \in {\bf N}$), so that
\begin{eqnarray}
  S = \sum_{i=0}^{N-2} \sum_{a \in \Ir} e_a(i) \, e_{a+1}(i+1) \; .
\end{eqnarray}
Obviously, $0 \leq S(\gamma) \leq N-1$  and $S(\gamma)$ is the
length of the path corresponding to the curve $\gamma$. The
curve given by $\gamma_{min}(i) := a$ (for some fixed $a \in \Ir$) is
a minimum, $\gamma_{max}(i) := i+a$ is a maximum of $S$.
We find
\begin{eqnarray}
  dS = \sum_{a \in \Ir} \lbrack e_{a,a+1}(N-1) - e_{a,a+1}(0)
       \rbrack
\end{eqnarray}
and therefore
\begin{eqnarray}
  dS(\gamma, \alpha) &=& \delta_{\alpha(N-1),\gamma(N-1)+1}
                         - \delta_{\alpha(0),\gamma(0)+1}  \\
  dS(\alpha, \gamma) &=& \delta_{\alpha(N-1),\gamma(N-1)-1}
                         - \delta_{\alpha(0),\gamma(0)-1}  \; .
\end{eqnarray}
For $\gamma_{min}$ this leads indeed to $dS(\gamma_{min}, \alpha)
\geq 0$ and $dS(\alpha, \gamma_{min}) \leq 0$.
\vskip.3cm
Admittedly, this example is too simple to be of real interest. It
nicely demonstrates, however, how our calculus works.
                                         \hfill {\Large $\Box$}

\section{Discrete differential manifolds and topological spaces}
\setcounter{equation}{0}
Let $\Omega({\cal M})$ be a differential calculus on a discrete set
$\cal M$ with elements $i,j, \ldots$ and $\lbrace e_I \rbrace_{I=
(i_1 i_2 \ldots)}$ a basis of $\Omega({\cal M})$ (as a $\Cx$-vector
space) consisting of basic $r$-forms. We represent these forms as
vertices of a digraph in such a way that vertices corresponding to
$(r+1)$-forms are below those corresponding to $r$-forms. If in the
differential calculus some $e_J$ appears in the expression (\ref{de...})
for $de_I$, then we draw an edge between the vertices representing $e_J$
and $e_I$. The result is a {\em Hasse diagram} which determines a
topology in the following way \cite{Sork91}. Any vertex together with
all lower lying vertices which are connected to it forms an open set.
Together with the empty and the whole set, the open sets obtained in
this way define a topology.
\vskip.3cm

Although for each $i \in {\cal M}$ we obtain an open set containing $i$,
we do not have points in $\cal M$ lying in intersections of these sets.
This suggests to consider the {\em extended set} $\hat{\cal M}$, the
points of which correspond to the vertices of the Hasse diagram.
With the topology determined by the Hasse diagram, $\hat{\cal M}$
becomes a topological space, the {\em extended space}.
\vskip.3cm

Let $X$ be a countable set and $\tau$ a locally finite topology on it,
i.e. a collection of open sets such that
\begin{eqnarray}
                 U(a) := \bigcap_{a \in U \in \tau}  U
\end{eqnarray}
is open for each $a \in \hat{\cal M}$. Then
\begin{eqnarray}       \label{preorder}
   a \; \hookrightarrow \; b \quad  \Leftrightarrow \quad a \in U(b)
\end{eqnarray}
defines a {\em preorder} (a transitive and reflexive relation) on
$X$ (cf \cite{Sork91}). This order relation is displayed in a Hasse
diagram in such a way that $a \hookrightarrow b$ iff the vertex $a$ is
connected from below to the vertex $b$.
\vskip.3cm

Now the following question arises. Is every finite
topological space (or, more generally, countable space with locally
finite topology) $(X, \tau)$ the extended space of some discrete
differential manifold ?
\vskip.3cm
\noindent
{\bf Definition.} A topological space $(X, \tau)$ is {\em generated} by
a discrete differential manifold $({\cal M}, \Omega({\cal M}))$ if \\
(1) $X$ is the extended set $\hat{\cal M}$ of $\cal M$, \\
(2) $\Omega({\cal M})$ induces the topology $\tau$ on $X$.
\vskip.3cm
\noindent
Our construction of topological (extended) spaces from differential
calculi on countable sets reaches many examples. Trivially, any
set with the discrete topology is `generated' (with $\Omega^r({\cal M})
= 0$ for $r > 0$). In general, the answer to the above question is
`no', however. A simple counterexample is the 2-point set with the
indiscrete topology (consisting of the empty and the whole set only).
This space is not of much interest, however. It is excluded if we
confine our considerations to $T_0$-spaces. $(X, \tau)$ is a
$T_0$-space if for each pair of distinct points in $X$ there is an open
set containing one point but not the other. This is the case if and
only if $\hookrightarrow$ is a {\em partial order} in which case $X$
receives the structure of a {\em poset} (partially ordered set)
\cite{Sork91}. But there are also counterexamples which are
$T_0$-spaces. On the 2-point set with elements $a$ and $b$ we may choose
as open sets $\lbrace a \rbrace, \, \lbrace a,b \rbrace$ (together
with the empty set). A $T_0$-counterexample with a 3-point set is
$ \bullet \hookrightarrow \bullet \hookleftarrow \bullet$.
One might think of imposing a stronger condition than $T_0$.
$T_1$ requires that each set which consists of a single point is
closed. This is too strong since the lowest points in a Hasse
diagram form open sets. The Hausdorff property $T_2$ is obviously
too strong.
\vskip.3cm
\noindent
{\bf Example.} Let $(X, \tau)$ be the topological 3-point space
determined by $ \bullet \hookleftarrow \bullet \hookrightarrow
\bullet $.
Let $\cal M$ be the 2-point set consisting of the first and the
last point. With the differential calculus on $\cal M$ corresponding
to the digraph $ \bullet \rightarrow \bullet $ (or $ \bullet
\leftarrow \bullet $) one finds that $(X, \tau)$ is `generated'. The
topology is $T_0$ but not $T_1$.
                                           \hfill {\Large $\Box$}

\vskip.3cm
\noindent
The condition for a topological space to be `generated' seems to
eliminate less useful topologies. It has still to be explored
how restrictive this condition actually is. The most interesting
aspect of a generated topological space is that all the information
about this space is already contained in a subset with a digraph
structure. In some cases this subset is much smaller than the
original set. It may be finite even when the original set is
infinite.

\section{Differentiability implies continuity}
\setcounter{equation}{0}
We have seen that a discrete differential manifold generates a
topological space. One should then expect that a differentiable map
between discrete differential manifolds extends to a continuous
map between the corresponding topological spaces.
\vskip.3cm

In this section we inessentially depart from our definition of the
Hasse diagram in the previous section. If there is a form which
is annihilated by $d$, then we draw a line to an additional lower lying
vertex which represents $0 \in \Omega({\cal M})$. In the topology
determined by the Hasse diagram this vertex stands for the empty set.
Instead of labeling the vertices of the Hasse diagram
by the elements of a basis of $\Omega({\cal M})$ consisting of basic
forms, it seems to be more appropriate to use the dual basis. The
reason is the following.
\vskip.3cm

The elements of $\cal M$ may be identified with linear maps dual
to the functions $e_i$. This suggests the construction of
an extended space $\hat{\cal M}$, the points of which are objects
dual to the forms $\lbrace e_I \rbrace$. In any case, the points
of $\hat{\cal M}$ correspond to the vertices of the Hasse diagram
(as in section VIII).
\vskip.3cm

If a map $\phi \, : \, {\cal M} \rightarrow {\cal M}'$ is
differentiable (with respect to differential calculi $\Omega({\cal M})$
and $\Omega({\cal M}')$), we will see that there is a natural extension
to a map $\phi_\star \, : \, \hat{\cal M} \rightarrow \hat{\cal M}'$
and this map is then continuous with respect to the topologies
defined by the Hasse diagrams (derived from $\Omega({\cal M})$ and
$\Omega({\cal M}')$, respectively).
\vskip.3cm
\noindent
Let $\Omega({\cal M})^\ast$ be the dual of $\Omega({\cal M})$ as
a $\Cx$-vector space. A basis $\lbrace e^J \rbrace$ is defined by
\begin{eqnarray}
    \delta^J_I = e^J (e_I) =: \langle e^J , e_I \rangle   \; .
\end{eqnarray}
These basis elements together with $0$ constitute the points of
$\hat{\cal M}$.
\vskip.3cm

A boundary operator $\partial \, : \, \Omega^r({\cal M})^\ast
\rightarrow \Omega^{r-1}({\cal M})^\ast$ dual to $d$ is now determined
via
\begin{eqnarray}                    \label{partial-d}
 \langle \partial e^J , e_I \rangle = \langle e^J , d e_I \rangle
 \; .
\end{eqnarray}
If $\phi \, : \, {\cal M} \rightarrow {\cal M}'$ is a differentiable
map with respect to differential calculi $\Omega({\cal M})$ and
$\Omega({\cal M}')$, so that $\phi^\star \, : \, \Omega({\cal M}')
\rightarrow \Omega({\cal M})$ is a homomorphism of differential
algebras, then
\begin{eqnarray}
   \langle e^J , \phi^\star e_{I'} \rangle
         = \langle \phi_\star e^J ,  e_{I'} \rangle
\end{eqnarray}
defines a linear map $\phi_\star \, : \, \Omega({\cal M})^\ast
\rightarrow \Omega({\cal M}')^\ast$.

\vskip.3cm
\noindent
{\bf Lemma.} If $\phi \, : \, {\cal M} \rightarrow {\cal M}'$ is
differentiable, then
\begin{eqnarray}
     \partial' \phi_\star = \phi_\star \partial   \; .
\end{eqnarray}
\vskip.1cm \noindent
{\bf Proof:}
\begin{eqnarray*}
    \langle \phi_\star \partial e^J , e_{I'} \rangle
   = \langle \partial e^J , \phi^\star e_{I'} \rangle
   = \langle e^J , d \, \phi^\star e_{I'} \rangle
   = \langle e^J , \phi^\star d' e_{I'} \rangle
   = \langle \phi_\star e^J , d' e_{I'} \rangle
   = \langle \partial' \phi_\star e^J , e_{I'} \rangle  \; .
\end{eqnarray*}
                               \hfill  {\Large $\Box$}

\vskip.3cm
\noindent
Let us write $e_J \in d e_I$ if $e_J$ appears in the expression
(\ref{de...}) for $de_I$. We also write $e^I \in \partial e^J$
if $e^I$ appears in the corresponding expression for $\partial e^J$.
{}From (\ref{partial-d}) we infer that
\begin{eqnarray}
     e_J \in d e_I \quad \Leftrightarrow \quad
     e^I \in \partial e^J  \; .
\end{eqnarray}
A simple consequence of the last Lemma is then
\begin{eqnarray}                       \label{phi_star-rel}
     e^I \in \partial e^J \quad \Rightarrow \quad
     \phi_\star e^I \in \partial' \phi_\star e^J    \; .
\end{eqnarray}
It follows from the next Lemma that $\phi_\star$ induces a map from
$\hat{\cal M}$ to $\hat{\cal M}'$. The presence of the auxiliary point
in $\hat{\cal M}'$ which represents $0 \in \Omega({\cal M}')^\ast$ and
stands for the empty set in $\hat{\cal M}'$ is necessary, since
in general $\phi_\star$ maps some dual forms to $0$. If $0$ were not
represented by a point in $\hat{\cal M}'$ then $\phi_\star$ would not
define a map $\hat{\cal M} \rightarrow \hat{\cal M}'$.

\vskip.3cm
\noindent
{\bf Lemma.} If $\phi \, : \, {\cal M} \rightarrow {\cal M}'$
is differentiable, then
\begin{eqnarray}
     \phi_\star \, e^J  =  e^{\phi(J)}
\end{eqnarray}
where $\phi(J) = \phi (j_1 \ldots j_r) := (\phi(j_1) \ldots
\phi(j_r))$.
\vskip.1cm \noindent
{\bf Proof:}
\begin{eqnarray*}
      \langle \phi_\star e^J , e_{I'} \rangle
   =  \langle e^J , \phi^\star e_{I'} \rangle
   =  \langle e^J , \sum_{I \in \phi^{-1}(I')} e_{I} \rangle
   =  \sum_{I \in \phi^{-1}(I')} \delta^J_I
   =  \delta^{\phi(J)}_{I'}
   =  \langle e^{\phi(J)} , e_{I'} \rangle     \; .
\end{eqnarray*}
                                   \hfill  {\Large $\Box$}

\vskip.3cm
\noindent
{\bf Proposition.} If $\phi \, : \, {\cal M} \rightarrow {\cal M}'$
is differentiable, then $\phi_\star$ is continuous as a map from
$\hat{\cal M}$ to $\hat{\cal M}'$ (with respect to the topologies
derived from the differential calculi on $\Omega({\cal M})$ and
$\Omega({\cal M}')$, respectively).
\vskip.1cm \noindent
{\bf Proof:}
Let $U$ be a nonempty open set in $\hat{\cal M}'$. Let us assume that
$\phi_\star^{-1}(U)$ is not open in $\hat{\cal M}$. In the Hasse
diagram defining the topology of $\hat{\cal M}$ there must then be
a vertex corresponding to a point in $\phi_\star^{-1}(U)$ which is
connected below to a vertex which does not correspond to a point
in $\phi_\star^{-1}(U)$. The latter vertex cannot be the one which
stands for the empty set in $\hat{\cal M}$ since $\phi_\star$ maps
it to the empty set in $\hat{\cal M}'$ which belongs to every open
set, so also to $U$. Hence there are dual forms $e^I, e^J$ such that
\begin{eqnarray*}
  \phi_\star e^I \in U \, , \quad \phi_\star e^J \not\in U
  \, , \quad e^I \in \partial e^J \; .
\end{eqnarray*}
But (\ref{phi_star-rel}) then implies
\begin{eqnarray*}
   \phi_\star e^I \in \partial' \phi_\star e^J  \; .
\end{eqnarray*}
Now we have a contradiction since $U$ is open.
                                         \hfill  {\Large $\Box$}

\section{Conclusions}
\setcounter{equation}{0}
Motivated by the results in \cite{DMH94} we have introduced the
notion of a `discrete differential manifold' and suggested to
regard it as an analogue of (continuous) differentiable manifolds.
\vskip.3cm

This structure has already been shown to be useful for lattice
field theories \cite{DMHS93}. As we have demonstrated in the
present work, it is also a convenient mathematical framework
to study mechanics on discrete spaces. We are, however, not
bound to an interpretation of the underlying discrete set as an
analogue of space or space-time. Let us give some other examples.
\vskip.3cm

The set ${\cal T}({\cal M})$ of all topologies on a given finite
set $\cal M$ is partially ordered by set inclusion. For
$\alpha, \beta \in {\cal T}({\cal M})$, the relation $\alpha
\subset \beta$ means that the topology $\alpha$ is coarser (weaker)
than $\beta$ or, equivalently, $\beta$ is finer (stronger) than
$\alpha$. There is a natural way to associate a graph with
${\cal T}({\cal M})$. We represent each topology by a vertex.
If $\alpha \subset \beta$ such that $\alpha \neq \beta$ and there is
no $\gamma \neq \alpha, \beta$ with $\alpha \subset \gamma \subset
\beta$, then we draw an edge between the vertex $\alpha$ and the vertex
$\beta$. Turning edges into arrows (or pairs of antiparallel arrows)
yields a digraph which defines a differential calculus on
${\cal T}({\cal M})$. A (differentiable) curve in ${\cal T}({\cal M})$
describes topology change on $\cal M$.
If, for example, we choose the orientation of the arrow from $\alpha$
to $\beta$ according to the relation $\alpha \subset \beta$, a
differentiable curve describes topology change only from coarser to
finer topology. It goes the other way if we reverse all the arrows.
See also \cite{Isha89} in this context.
\vskip.3cm

Any differential calculus on a discrete set $\cal M$ is the quotient
of the universal differential calculus by some differential ideal
$\cal I$. The inclusion of ideals then partially orders the set
${\cal DC}({\cal M})$ of all differential calculi on $\cal M$, i.e.
the set of all discrete differential manifolds with point space
$\cal M$. ${\cal DC}({\cal M})$ naturally carries the structure of
a graph. If the relation ${\cal I} \subset {\cal I'}$ holds with
${\cal I} \neq {\cal I'}$ and there is no ${\cal I''} \neq {\cal I},
{\cal I'}$ such that ${\cal I} \subset {\cal I''} \subset {\cal I'}$,
then we draw an edge between the vertices representing the two
differential calculi $\tilde{\Omega}({\cal M})/{\cal I}$ and
$\tilde{\Omega}({\cal M})/{\cal I'}$. Turning edges into arrows
(or pairs), we obtain a digraph which then defines a differential
calculus on ${\cal DC}({\cal M})$. A curve in ${\cal DC}({\cal M})$
describes a change of the differential calculus on $\cal M$.
\vskip.3cm

An algebraic approach to discrete mechanics appeared recently
in \cite{Baez+Gill94}. The authors of that paper considered a
commutative algebra $\cal A$ over a commutative ring $k$. It is
assumed that $\cal A$ is freely generated, say, by elements
$x_1, \ldots, x_n$. It is then possible to have functions commuting
with differentials (`K\"ahler differentials'). Choosing $k = \Ir$, for
example, the formalism is able to describe motion on the lattice
$\Ir^n$. In contrast, we consider algebras of functions over $k =
\Cx$. On a discrete set these are subject to the constraints
(\ref{e-rels}) which force us to work with a `noncommutative
differential calculus'.
\vskip.3cm

A next step in our programme should be a formulation of discrete
quantum mechanics (see \cite{Zakh93} and references given there)
on discrete differential manifolds. Here a path integral approach
is prefered.
\vskip.3cm

The association of a digraph with a discrete differential manifold
suggests a natural way how to quantize it, namely to turn it into
a `quantum network' (e.g., in the sense of Finkelstein \cite{Fink89},
see also \cite{Anto94}). This is a further interesting route to
proceed.

\acknowledgments
We have to thank John Madore for some helpful discussions.

\end{document}